\documentclass[twocolumn,aps,prc,amsfonts,preprintnumbers,superscriptaddress,showpacs,nofootinbib]{revtex4}
\pdfoutput=1
\usepackage{graphics}
\usepackage{amsmath}
\usepackage{amssymb}
\usepackage{psfrag}
\usepackage{epsfig}
\usepackage{epsf}
\usepackage{float}
\usepackage{slashed}

\usepackage{physics}
\usepackage{tensor}

\allowdisplaybreaks

\newcommand{\beq}{\begin{equation}}
\newcommand{\eeq}{\end{equation}}
\newcommand{\beqa}{\begin{eqnarray}}
\newcommand{\eeqa}{\end{eqnarray}}

\begin{document}

\title{Scattering phase shifts and mixing angles for an arbitrary number
  of coupled channels on the lattice}

\author{Lukas Bovermann}
\email[]{lukas.bovermann@rub.de}
\affiliation{Ruhr-Universit\"at Bochum, Fakult\"at f\"ur Physik und Astronomie,
Institut f\"ur Theoretische Physik II, 
  D-44780 Bochum, Germany}
\author{Evgeny Epelbaum}
\email[]{evgeny.epelbaum@rub.de}
\affiliation{Ruhr-Universit\"at Bochum, Fakult\"at f\"ur Physik und Astronomie,
Institut f\"ur Theoretische Physik II, 
  D-44780 Bochum, Germany}
\author{Hermann Krebs}
\email[]{hermann.krebs@rub.de}
\affiliation{Ruhr-Universit\"at Bochum, Fakult\"at f\"ur Physik und Astronomie,
Institut f\"ur Theoretische Physik II, 
  D-44780 Bochum, Germany}
\author{Dean Lee}
\email[]{leed@frib.msu.edu}
\affiliation{Facility for Rare Isotope Beams and Department of Physics
  and Astronomy, Michigan State University, Michigan 48824, USA}
\date{December 11, 2019}

\begin{abstract}
We present a lattice method for determining scattering phase shifts and mixing
angles for the case of an arbitrary number of coupled
channels. Previous nuclear lattice effective field theory simulations were restricted to mixing of up to
two partial waves for scattering of two spin-$1/2$ particles, which is
insufficient for analyzing 
nucleon-nucleus or nucleus-nucleus scattering processes. 
In the proposed method, the phase shifts and mixing angles are extracted from the radial wave
functions obtained by projecting the three-dimensional lattice
Hamiltonian onto the partial wave basis. We use a spherical wall potential
as a boundary condition along with a channel-mixing auxiliary potential to
construct the full-rank $S$~matrix. Our method can be applied to particles
with any spin, but we focus here on scattering of two spin-$1$
bosons involving up to four coupled channels. For a
considered test potential, the phase shifts and mixing angles
extracted on the lattice are shown to agree with the ones
calculated by solving the Schr\"odinger equation in the continuum. 
\end{abstract}


\maketitle

\section{Introduction}

Lattice simulations provide a powerful computational approach to
systems of strongly interacting particles, which is widely used in
condensed matter, nuclear, and particle physics. In particular, lattice
gauge theory is the only known numerical method that allows one to
directly solve QCD in the nonperturbative domain. Here, remarkable
progress has been achieved in the recent decades due to the rapid
increase of computational power and algorithmic efficiency. In particular,
high-precision lattice QCD calculations of hadronic observables, such as
the masses and decay constants, are already available for physical
values of the quark masses \cite{Tanabashi:2018oca}. While hadronic
reactions and resonance properties can also be addressed in lattice
QCD, such calculations appear to be much more challenging and require
developing reliable methods for relating the scattering amplitude to 
discrete finite-volume spectra accessible in lattice simulations, see
Refs.~\cite{Lage:2009zv,Bernard:2010fp,Hansen:2012tf,Briceno:2012rv,Hammer:2017uqm,Hammer:2017kms,Briceno:2017tce}
for recent work 
along this line and 
\cite{Briceno:2017max} for a review article.   

Lattice methods have also proven to be very efficient in describing
low-energy nuclear systems in the framework of chiral effective field theory
(EFT). Recently, the chiral expansion of the nucleon-nucleon (NN) potential
has been pushed to fifth order (N$^4$LO)
\cite{Epelbaum:2014efa,Epelbaum:2014sza,Entem:2014msa,Entem:2017gor}
within the continuum 
formulation. The NN potentials derived in chiral EFT in Ref.~\cite{Reinert:2017usi}
allow, for the first time, for a nearly perfect description of the
neutron-proton and proton-proton scattering data below the pion
production threshold, which is comparable to or even better than that
based on the available phenomenological potentials. Three- and
four-nucleon forces have been worked out completely up to fourth order
(N$^3$LO) of the chiral expansion
\cite{Ishikawa:2007zz,Bernard:2007sp,Bernard:2011zr,Epelbaum:2007us},
see also
Refs.~\cite{Krebs:2012yv,Krebs:2013kha,Epelbaum:2014sea,Girlanda:2011fh}
for the derivation of selected contributions at N$^4$LO and
Refs.~\cite{Epelbaum:2008ga,Epelbaum:2012vx,Machleidt:2011zz} for
review articles. To apply the interactions derived in chiral EFT
to few- and many-nucleon systems, it is necessary to solve  the quantum
mechanical $A$-body problem, which can be achieved using continuum
\textit{ab initio} methods including Faddeev-Yakubovsky equations \cite{Gloeckle:1995jg}, the no-core
configuration interaction approach \cite{Barrett:2013nh},
coupled-cluster expansions \cite{Hagen:2012fb}, the in-medium similarity
renormalization group approach \cite{Hergert:2012nb}, self-consistent Green's
function \cite{Soma:2012zd}, or quantum Monte Carlo methods
\cite{Lovato:2013cua}. Alternatively,
a discretized version of chiral EFT \cite{Lee:2008fa,Lee:2016fhn,UGMlectures} has been
successfully applied to a
broad range of nuclear systems. This approach has an appealing feature
of being well suited for dealing with strongly clustered systems such
as the famous Hoyle state in $^{12}$C
\cite{Epelbaum:2011md,Epelbaum:2012qn,Epelbaum:2013paa} and some of
the low-lying states of  $^{16}$O,
which often represent a challenge for continuum methods. See 
Ref.~\cite{Freer:2017gip} for a recent review on clustering in
light nuclei. So far, nuclear
lattice simulations have been carried out for light- and medium-mass
nuclei and neutron matter up to third order  in the chiral
expansion
\cite{Borasoy:2007vk,Epelbaum:2009zsa,Lahde:2013uqa}. For a recent
lattice EFT study of NN scattering at N$^3$LO see Ref.~\cite{Li:2018ymw}. This method was
also employed to study the dependence of the triple-$\alpha$ process on the
fundamental constants of nature
\cite{Epelbaum:2012iu,Epelbaum:2013wla}, see Ref.~\cite{Meissner:2014pma} for a
related discussion, to investigate  
the isotopic dependence of nuclear clustering
\cite{Elhatisari:2017eno} and to determine the features of the nuclear force essential
for nuclear binding \cite{Lu:2018bat}.
It is important to emphasize
that the development of  chiral EFT interactions is more
difficult on the lattice than in the continuum as it requires establishing
efficient techniques for extracting the scattering amplitude from the
finite-volume discrete spectra and for dealing with the breaking of
rotational \cite{Alarcon:2017zcv,Klein:2018iqa} and Galilean invariance \cite{Li:2019ldq} due to
nonzero lattice spacing. L\"uscher's finite-volume method is one
possible approach to compute
scattering phase shifts on the lattice. For lattice QCD
applications, there have been many recent advances on coupled-channel
calculations and partial-wave mixing using L\"uscher's formalism
\cite{Briceno:2014oea, Moir:2016srx, Briceno:2017qmb, Woss:2018irj, Woss:2019hse}.
However, for lattice EFT calculations of heavier nuclear systems, the large nuclear
binding energies and very small finite-volume scattering energies
make it difficult to implement L\"uscher's method with accuracy.
For this reason, a more robust approach
based on the spherical wall boundary conditions \cite{Carlson:1984zz} was used in
Refs.~\cite{Borasoy:2007vy,Lu:2015riz,Alarcon:2017zcv,Li:2018ymw}. This technique is
not only applicable to calculations of  NN 
phase shifts on the lattice, but can also be combined with the
adiabatic projection method \cite{Rokash:2015hra,Elhatisari:2016hby}, which allows one to access nuclear
reactions via lattice simulations, see Ref.~\cite{Elhatisari:2015iga} for the first
\textit{ab initio} study of $\alpha$-$\alpha$ scattering. However, the spherical wall
method has so far only been applied to uncoupled partial waves and
the cases of two coupled channels, which is insufficient
for studying nuclear reactions. The purpose of this paper
is to generalize this technique to an arbitrary number of coupled
channels. 

Our paper is organized as follows. In Sec.~\ref{sec2}, we introduce
the lattice notation, review the method of Ref.~\cite{Lu:2015riz} to compute the
scattering parameters with up to two coupled channels, and extend this
approach to scattering of particles or nuclear clusters of an
arbitrary high spin. As an application, we consider in Sec.~\ref{sec3}
the scattering problem of two spin-$1$ bosons using a toy-model
potential, which is similar to the one from
Refs.~\cite{Borasoy:2007vy,Lu:2015riz}. The main results of our study are
summarized in Sec.~\ref{sec4}.

\section{Scattering of two particles with arbitrary spin on the lattice}
\label{sec2}

\subsection{Calculational setup}

We employ a periodic cubic lattice with the
length $L$ and spacing $a$,
and define orthonormal lattice states $\ket{\mathbf{r}}$ with
\begin{align}
\nonumber
&r_1, r_2, r_3 = 0, \dots, L-1,
\end{align}
\begin{align}
&\ket{\mathbf{r}} = \ket{\mathbf{r} + L \mathbf{\hat{e}}_1}
= \ket{\mathbf{r} + L \mathbf{\hat{e}}_2} = \ket{\mathbf{r} + L \mathbf{\hat{e}}_3}
\end{align}
due to the periodic boundary condition. All quantities in this section
are given in dimensionless lattice units, i.e.,~they must be
multiplied by an appropriate power of the lattice spacing $a$ to
obtain their physical values. In the following, we briefly review the
method to compute the scattering phase shifts and mixing angles for up
to  two coupled channels introduced in
Ref.~\cite{Lu:2015riz}, which will then be generalized to the case of
three or more coupled channels.

We consider the general scattering problem of two particles with spins $s_1, s_2$
and masses $m_1, m_2$ interacting with the potential $V(\mathbf{r})$.
The free Hamiltonian in the center-of-mass (c.m.) system is discretized
as \cite{Borasoy:2007vy}
\begin{align}
\nonumber
H_0 \ket{\mathbf{r}} &= \frac{49}{12 \mu} \ket{\mathbf{r}}
                                -\frac{3}{4 \mu} \sum_{i=1}^3 \left(
                                \ket{\mathbf{r} + \mathbf{\hat{e}}_i} +
                                \ket{\mathbf{r} - \mathbf{\hat{e}}_i} \right) \\
                                \nonumber
                                &\qquad +\frac{3}{40 \mu} \sum_{i=1}^3 \left(
                                \ket{\mathbf{r} + 2\mathbf{\hat{e}}_i} +
                                \ket{\mathbf{r} - 2\mathbf{\hat{e}}_i} \right) \\ 
&\qquad -\frac{1}{180 \mu} \sum_{i=1}^3 \left( \ket{\mathbf{r} + 3\mathbf{\hat{e}}_i} + \ket{\mathbf{r} - 3\mathbf{\hat{e}}_i} \right)
\end{align}
using the reduced mass $\mu = m_1 m_2 / (m_1 + m_2)$. The above
expression corresponds to the $O(a^4)$-improved free lattice
Hamiltonian. 
To avoid artifacts induced by the periodic boundary conditions, it is
convenient to use a spherical wall boundary condition by adding the potential
\begin{align}
V_\mathrm{wall}(\mathbf{r}) = \Lambda \theta(r - R_W),
\end{align}
where $\theta$ is the Heaviside function, $R_W$ is the wall
radius, and $\Lambda$ is a large positive constant\footnote{Following
  Ref.~\cite{Lu:2015riz}, we use the value $\Lambda = 10^6$ (given in dimensionless lattice units) in the
  numerical calculations.}
\cite{Borasoy:2007vy}. Calculating the scattering parameters at low momenta
usually requires large lattices, which makes the analysis
computationally expensive. It is more convenient to introduce an
auxiliary potential outside of the range of $V$, which can be chosen, e.g.,~of a Gaussian type 
\begin{align}
V_\mathrm{aux}(\mathbf{r}) = V_0 \exp[-(r - R_W)^2] \,
  \theta(R_W - r) 
\end{align}
with $V_0 \leq 0$  in order to control the eigenenergies of the
Hamiltonian \cite{Lu:2015riz}. The complete Hamiltonian including all
contributions is then given by
\begin{align}
H = H_0 + V + V_\mathrm{wall} + V_\mathrm{aux} \, .
\end{align}

\subsection{Projection onto partial waves}

The three-dimensional problem  can be
reduced to the one-dimensional one by 
defining radial states for a partial wave
$^{2s+1} l_j$,
\begin{align}
\nonumber
\ket{R}_{s,l,j} &= \sum_{\mathbf{r}} \sum_{l_z, s_z} \sum_{s_{1,z}} \sum_{s_{2,z}}
C^{j,l,s}_{j_z,l_z,s_z} C^{s,s_1,s_2}_{s_z,s_{1,z},s_{2,z}} \\
                    &\qquad \times Y_{l,l_z} (\mathbf{\hat{r}}) \delta_{r,R}
                    \ket{\mathbf{r}} \otimes
                    \ket{s_{1,z},s_{2,z}}, 
\end{align}
i.e.,~the lattice sites $\mathbf{r}$ with the same radial distance $R$ are
grouped together according to the irreducible representations of the
rotational group. Here, $C^{j,l,s}_{j_z,l_z,s_z}$
and $C^{s,s_1,s_2}_{s_z,s_{1,z},s_{2,z}}$ are the
Clebsch-Gordan coefficients for the spin-orbit and spin-spin couplings,
respectively. The spherical harmonics $Y_{l,l_z}$ behave like
$Y_{l,l_z} (\mathbf{0}) = \delta_{l,0} / \sqrt{4\pi}$
at the origin. Since the results obtained here do not depend on
$j_z$ in the continuum limit, we can choose $j_z = 0$. 

Note that the radial states have to be normalized by dividing them by the square root of their norm. States that are not linearly independent or have vanishing norm must be omitted to make the norm matrix invertible. Afterwards, the Hamiltonian can be projected onto the normalized radial states. For $n$ coupled channels with
\begin{align}
\ket{R}_\alpha := \ket{R}_{s_\alpha, l_\alpha, j_\alpha}
\text{ for }
\alpha = 1, \dots, n,
\end{align}
one has
\begin{align}
\nonumber
[H_R (R_1, R_2)]_{\alpha \beta}
&= \sum_{\alpha^\prime, \beta^\prime = 1}^n
[N^{-1/2} (R_1)]_{\alpha \alpha^\prime}
[N^{-1/2} (R_2)]_{\beta^\prime \beta} \\
&\qquad \times \tensor[_{\alpha^\prime}]{\bra{R_1} H \ket{R_2}}{_{\beta^\prime}}
\,,
\end{align}
where $N^{-1/2}$ is the inverse square root of the norm matrix
\begin{align}
[N(R)]_{\alpha \alpha^\prime}
= \tensor[_{\alpha}]{\braket{R}{R}}{_{\alpha^\prime}}.
\end{align}
Multiplying the eigenvectors of the projected Hamiltonian $H_R$ by $N^{-1/2}$ from the left yields the radial wave functions
\begin{align}
\psi(r) = [\psi_1(r), \dots, \psi_n(r)]^T.
\end{align}

\subsection{Single-channel case}

Outside of the range of the potential, the radial wave functions are
linear combinations of the spherical Hankel functions $h_l^\pm(pr)$,
where $p$ is the momentum in the center-of-mass system. For a single scattering channel, one obtains a wave function of the form
\begin{align}
\psi(r) = A h_l^-(pr) + B h_l^+(pr),
\end{align}
which allows one to extract the phase shift $\delta_l$ from the
$S$~matrix via
\begin{align}
S = B/A = e^{2i\delta_l}.
\end{align}
The coefficients $A$, $B$ are computed by fitting the spherical Hankel
functions to the wave function in an interval $[R_I,
R_O]$ outside of the range of the potential. 
The momentum $p$ is
determined from the eigenenergy of the Hamiltonian using the lattice
dispersion relation 
\begin{align}
\nonumber
E(\mathbf{p}) &= \frac{49}{12\mu} - \frac{3}{2\mu} \sum_{i=1}^3 \cos(p_i) + \frac{3}{20\mu} \sum_{i=1}^3 \cos(2 p_i) \\
&\qquad - \frac{1}{90\mu} \sum_{i=1}^3 \cos(3 p_i).
\end{align}
This equation can be expressed in spherical coordinates with
\begin{align}
\mathbf{p} = (p \sin \theta \cos \phi, \; p \sin \theta \sin \phi, \; p \cos \theta).
\end{align}
In order to remove the angular dependence, the dispersion relation must be projected onto partial waves as well:
\begin{align}
\nonumber
&E_{s,l,j} (p) = \int d \Omega_p
\sum_{l_z, s_z} \sum_{l_z^\prime, s_z^\prime}
\sum_{s_{1,z}} \sum_{s_{2,z}}
C^{j,l,s}_{0,l_z,s_z}
C^{s,s_1,s_2}_{s_z,s_{1,z},s_{2,z}} \\
&\qquad \times C^{j,l,s}_{0,l_z^\prime,s_z^\prime}
C^{s,s_1,s_2}_{s_z^\prime,s_{1,z},s_{2,z}}
Y_{l,l_z}^* (\mathbf{\hat{p}}) Y_{l,l_z^\prime} (\mathbf{\hat{p}}) E(\mathbf{p}) .
\end{align}
The angular integration can be facilitated by Taylor-expanding
$E(\mathbf{p})$
up to order $O(p^\kappa)$, which should be sufficiently high to yield
accurate results up to the cutoff momentum $\pi/a$. 
\footnote{In numerical
  calculations, we use $\kappa = 30$.} Afterward, the c.m. system momentum can be computed by solving $E_{s,l,j}(p)$ for $p$.

\subsection{Scattering with two coupled channels}

For two coupled channels, the $S$~matrix must be constructed as
\begin{align}
S = \left( \begin{array}{ll}
v_1^+ & v_2^+
\end{array} \right)
\left( \begin{array}{ll}
v_1^- & v_2^-
\end{array} \right)^{-1},
\end{align}
where $v_1^\pm, v_2^\pm$ are linearly independent
two-component vectors containing the coefficients in front of the
spherical Hankel functions $h_{l_i}^\pm$. A simple way to obtain these
coefficients would be to extract them from a complex wave function 
\begin{align}
\psi(r) = \left( \begin{array}{l}
A_1 h_{l_1}^-(pr) + B_1 h_{l_1}^+(pr) \\
A_2 h_{l_1}^-(pr) + B_2 h_{l_2}^+(pr)
\end{array} \right)
\end{align}
and its complex conjugate
\begin{align}
\psi^*(r) = \left( \begin{array}{l}
A_1^* h_{l_1}^+(pr) + B_1^* h_{l_1}^-(pr) \\
A_2^* h_{l_1}^+(pr) + B_2^* h_{l_2}^-(pr)
\end{array} \right)
\end{align}
using
\begin{align}
\nonumber
&v_1^- = (A_1, A_2)^T, & &v_2^- = (B_1^*, B_2^*)^T, \\
&v_1^+ = (B_1, B_2)^T, & &v_2^+ = (A_1^*, A_2^*)^T.
\end{align}
However, the Hamiltonian $H_R$ commutes with the time-reversal operator $T$ so that it holds
\begin{align}
\psi^* = T \psi = \psi
\quad \Rightarrow \quad
v_1^\pm = v_2^\pm,
\end{align}
i.e.,~the vectors $v_1^\pm, v_2^\pm$ are linearly
dependent. Thus, one runs into the problem of having only one
independent solution per lattice energy. In order to circumvent this 
problem, an auxiliary imaginary potential term that breaks the time-reversal symmetry can be added to the
Hamiltonian:
\begin{align}
(H_R + U) \psi(r) = E \psi(r)
\label{equ:SEQWithMixPot}
\end{align}
with
\begin{align}
U(r) = U_0 \delta_{r,R_M} \left( \begin{array}{rr}
0 & i \\
-i & 0
\end{array} \right)
\label{equ:MixPot2ChImag}
\end{align}
and $U_0 \in \mathbb{R}$. The radius $R_M$
should lie outside the range of the test potential and can be chosen 
close to the spherical wall radius,  $R_M \lesssim R_W$.
Because the matrix in
Eq.~(\ref{equ:MixPot2ChImag}) mixes the two channels, $U(r)$ will be
referred to as the mixing potential. Finally, the $S$~matrix is
decomposed according to the Blatt-Biedenharn parametrization
\cite{Blatt:1952zza} 
\begin{align}
\nonumber
S &= \left( \begin{array}{rr}
\cos \epsilon & \sin \epsilon \\
-\sin \epsilon & \cos \epsilon
\end{array} \right)^{-1}
\left( \begin{array}{cc}
e^{2i \delta_1} & 0 \\
0 & e^{2i \delta_2}
\end{array} \right) \\
&\qquad \times \left( \begin{array}{rr}
\cos \epsilon & \sin \epsilon \\
-\sin \epsilon & \cos \epsilon
\end{array} \right)
\end{align}
with the phase shifts $\delta_1, \delta_2$ and the mixing angle
$\epsilon$. Since the lattice dispersion relation can yield slightly
different momenta $p_1$, $p_2$ for the two channels at the same
energy, we assume that the phase shift $\delta_\alpha$ is measured at
momentum $p_\alpha$ and that the mixing angle is measured at the
average momentum $(p_1 + p_2)/2$. 

\subsection{Scattering with an arbitrary number of coupled channels}

If $n > 2$ coupled channels must be considered, one needs $n$ linearly
independent wave functions in each channel. The complex conjugation is
not sufficient for this purpose because it can only generate two
independent solutions
$\psi$ and $\psi^*$. In order to find an alternative approach, we first
consider the case of two coupled partial waves again. The two-channel
wave function
\begin{align}
\psi(r) = \left[ \psi_1(r) , \; \psi_2(r) \right]^T
\end{align}
can be rewritten as
\beq
\psi^\prime(r) = \left[ 
\mathfrak{R} \, \psi_1(r), 
\mathfrak{I} \, \psi_1(r), 
\mathfrak{R} \, \psi_2(r), 
\mathfrak{I} \, \psi_2(r)
 \right]^T .
\label{equ:WaveFunc2ChReal}
\eeq
To reproduce Eqs.~(\ref{equ:SEQWithMixPot}) and
(\ref{equ:MixPot2ChImag}), the radial Hamiltonian and the mixing
potential must be modified accordingly:
\begin{align}
\nonumber
&H_R^\prime = \left( \begin{array}{cc|cc}
[H_R]_{11} & 0 & [H_R]_{12} & 0 \\
0 & [H_R]_{11} & 0 & [H_R]_{12} \\
\hline
[H_R]_{21} & 0 & [H_R]_{22} & 0 \\
0 & [H_R]_{21} & 0 & [H_R]_{22} \\
\end{array} \right), \\
&U^\prime = U_0 \; \delta_{r,R_M} \left( \begin{array}{rr|rr}
0 & 0 & 0 & -1 \\
0 & 0 & 1 & 0 \\
\hline
0 & 1 & 0 & 0 \\
-1 & 0 & 0 & 0
\end{array} \right) .
\label{equ:MixPot2ChReal}
\end{align}
On the other hand, instead of using Eq.~(\ref{equ:WaveFunc2ChReal}),
we can regard the wave function vector as having four independent complex components
\beq
\psi^\prime(r) = \left[
\psi_1^\prime(r), \;
\psi_2^\prime(r), \;
\psi_3^\prime(r), \;
\psi_4^\prime(r)
 \right]^T.
\label{equ:WaveFunc2ChComplex}
\eeq
A natural  extension of Eqs.~(\ref{equ:MixPot2ChReal}),~(\ref{equ:WaveFunc2ChComplex}) to three coupled
scattering channels is given by introducing 
\begin{widetext}
\begin{align*}
H_R^\prime = \left( \begin{array}{ccc|ccc|ccc}
[H_R]_{11} & 0 & 0 & [H_R]_{12} & 0 & 0 & [H_R]_{13} & 0 & 0 \\
0 & [H_R]_{11} & 0 & 0 & [H_R]_{12} & 0 & 0 & [H_R]_{13} & 0 \\
0 & 0 & [H_R]_{11} & 0 & 0 & [H_R]_{12} & 0 & 0 & [H_R]_{13} \\
\hline
[H_R]_{21} & 0 & 0 & [H_R]_{22} & 0 & 0 & [H_R]_{23} & 0 & 0 \\
0 & [H_R]_{21} & 0 & 0 & [H_R]_{22} & 0 & 0 & [H_R]_{23} & 0 \\
0 & 0 & [H_R]_{21} & 0 & 0 & [H_R]_{22} & 0 & 0 & [H_R]_{23} \\
\hline
[H_R]_{31} & 0 & 0 & [H_R]_{32} & 0 & 0 & [H_R]_{33} & 0 & 0 \\
0 & [H_R]_{31} & 0 & 0 & [H_R]_{32} & 0 & 0 & [H_R]_{33} & 0 \\
0 & 0 & [H_R]_{31} & 0 & 0 & [H_R]_{32} & 0 & 0 & [H_R]_{33}
\end{array} \right),
\end{align*}
\begin{align}
U^\prime = U_0 \; \delta_{r,R_M} \left( \begin{array}{rrr|rrr|rrr}
0 & 0 & 0 & 0 & -1 & 1 & 0 & 1 & -1 \\
0 & 0 & 0 & 1 & 0 & 1 & 1 & 0 & 1 \\
0 & 0 & 0 & 1 & 1 & 0 & 1 & 1 & 0 \\
\hline
0 & 1 & 1 & 0 & 0 & 0 & 0 & 1 & 1 \\
-1 & 0 & 1 & 0 & 0 & 0 & 1 & 0 & -1 \\
1 & 1 & 0 & 0 & 0 & 0 & 1 & 1 & 0 \\
\hline
0 & 1 & 1 & 0 & 1 & 1 & 0 & 0 & 0 \\
1 & 0 & 1 & 1 & 0 & 1 & 0 & 0 & 0 \\
-1 & 1 & 0 & 1 & -1 & 0 & 0 & 0 & 0 
\end{array} \right),
\quad \psi^\prime (r) = \left( \begin{array}{c}
\psi_1^\prime (r) \\
\psi_2^\prime (r) \\
\psi_3^\prime (r) \\
\psi_4^\prime (r) \\
\psi_5^\prime (r) \\
\psi_6^\prime (r) \\
\psi_7^\prime (r) \\
\psi_8^\prime (r) \\
\psi_9^\prime (r)
\end{array} \right).
\end{align}
\end{widetext}

A generalization to $n$ channels is straightforward:
\beqa
\nonumber
&[H_R^\prime]_{\alpha^\prime+(\alpha-1)n, \beta^\prime+(\beta-1)n}
= [H_R]_{\alpha, \beta} \delta_{\alpha^\prime, \beta^\prime}, \\
\nonumber
&U^\prime_{\alpha^\prime+(\alpha-1)n, \beta^\prime+(\beta-1)n}
= U_0 \delta_{r,R_M} (1 - \delta_{\alpha, \beta}) \\
& \qquad \qquad \qquad \times (1 - \delta_{\alpha^\prime, \beta^\prime} - 2 \delta_{\alpha, \alpha^\prime}
\delta_{\beta, \beta^\prime})
\label{temp1}
\eeqa
for $\alpha, \alpha^\prime, \beta, \beta^\prime = 1, \dots, n$. The wave function vector has the form
\begin{align}
\psi^\prime(r) = [\psi_1^\prime(r), \dots, \psi_{n^2}^\prime(r)]^T,
\end{align}
where $\psi_{\beta+(\alpha-1)n}^\prime$ denotes the $\beta$th wave
function for the $\alpha$th scattering channel with $\alpha, \beta =
1, \dots, n$. 

More generally, any Hermitian matrix that produces $n$ linearly
independent solutions in every channel can be used to define  the
mixing potential, i.e.,~it must hold ${U^\prime}^\dagger = U^\prime$
and the matrix $M$ with 
\begin{align}
M_{\alpha,\beta} = \sum_{\alpha^\prime, \beta^\prime = 1}^n [H_R^\prime + U^\prime]_{\beta+(\alpha-1)n,\beta^\prime+(\alpha^\prime-1)n} h_{l_{\alpha^\prime}}^\pm
\end{align}
must have rank $n$. The particular choice for the mixing potential in
Eq.~(\ref{temp1}) is consistent with the one employed for two channels in
Ref.~\cite{Lu:2015riz}. 

Each component of the wave function vector has the form
\beq
\psi_{\beta+(\alpha-1)n}^\prime(r) = A_{\alpha\beta} h_{l_\alpha}^-(pr) 
+ B_{\alpha\beta} h_{l_\alpha}^+(pr)
\eeq
with $\alpha, \beta = 1, \dots, n$. Since it holds
\begin{align}
\left( \begin{array}{c}
B_{1\beta} \\
\vdots \\
B_{n\beta}
\end{array} \right)
= S \left( \begin{array}{c}
A_{1\beta} \\
\vdots \\
A_{n\beta}
\end{array} \right)
\end{align}
for $\beta = 1, \dots, n$, one can construct the $S$~matrix as
\begin{align}
S=
\left( \begin{array}{ccc}
B_{11} & \cdots & B_{1n} \\
\vdots & \ddots & \vdots \\
B_{n1} & \cdots & B_{nn}
\end{array} \right)
\left( \begin{array}{ccc}
A_{11} & \cdots & A_{1n} \\
\vdots & \ddots & \vdots \\
A_{n1} & \cdots & A_{nn}
\end{array} \right)^{-1}.
\end{align}
The Blatt-Biedenharn parametrization also has to be extended to $n > 2$ coupled channels \cite{Blatt:1952zza}:
\begin{align}
S = O^{-1} \, \mathrm{diag}(e^{2i\delta_1}, \dots, e^{2i\delta_n}) \, O,
\end{align}
where $O$ is a real orthogonal matrix. (This decomposition is equivalent to computing the eigenvalues and eigenvectors of $S$.) Again, the phase shift $\delta_\alpha$ is assigned to the momentum $p_\alpha$ in scattering channel $\alpha$. For simplicity, we define the mixing angles as
\begin{align}
\epsilon_{\alpha\beta} \left( p = \frac{p_\alpha+p_\beta}{2} \right) = \tan^{-1} O_{\alpha\beta}
\label{equ:MixAngle}
\end{align}
for $\alpha, \beta = 1, \dots, n$ and $\beta > \alpha$ because a real orthogonal $n \times n$~matrix can be
given by $n (n-1) / 2$ real parameters.\footnote{Note that Eq.~(\ref{equ:MixAngle}) allows one to extract
the mixing angles from the matrix $O$, which is sufficient for the purpose of this paper. The equation
can, however, not be used to reconstruct the $S$~matrix from the phase shifts and mixing angles in a
unique way. This problem could be avoided by parametrizing the orthogonal matrix $O$ in terms of
generators of the rotation group.}

\section{Test case: scattering of two spin-$1$ particles}
\label{sec3}

The method described in Sec.~\ref{sec2} allows one to determine
scattering phase
shifts and mixing angles on the lattice for an arbitrary number of
coupled channels and for any type of particles. As a concrete example,
we consider the scattering problem  of two spin-$1$ bosons having nearly the
same mass as the deuteron,
$m_{1,2} = 2 m_N = 2 \times 938.92 \; \mathrm{MeV}$. As a test
potential, we employ the corresponding generalization of the toy-model
potential used for two spin-1/2 fermions in
Refs.~\cite{Borasoy:2007vy,Lu:2015riz}: 
\beq
V(\mathbf{r}) = C \left( 1 + \frac{s_{12}(\mathbf{r})}{r_0^2} \right) \exp \left( -\frac{r^2}{2r_0^2} \right),
\eeq
where the spin-dependent part is given by
\beq
s_{12}(\mathbf{r}) = 3 (\mathbf{r} \cdot \mathbf{s}_1) (\mathbf{r} \cdot \mathbf{s}_2) - (\mathbf{s}_1 \cdot \mathbf{s}_2) r^2.
\eeq
Here, $\mathbf{s}_1$ and $\mathbf{s}_2$ denote the spin matrices for the
considered particles. The constants are set to $C = -2 \; \mathrm{MeV}$ and $r_0
= 0.02 \; \mathrm{MeV}^{-1}$, and the lattice spacing is chosen to be $a = (100 \;
\mathrm{MeV})^{-1} = 1.9733 \; \mathrm{fm}$.  
Notice that by projecting the test potential onto
partial waves one obtains up to four coupled scattering channels. We calculate the phase shifts
and mixing angles for the following cases:
\begin{itemize}
\item[(i)] Uncoupled channels: $^3 P_0$,
  $^3 P_1$, $^3 D_2$,
  $^3 F_3$, $^3 G_4$,
  $^3 H_5$, $^5 D_1$;
\item[(ii)] Two coupled channels: $^3 SD_1$, $^3 PF_2$, $^3 DG_3$, $^3 FH_4$, $^1 S_0 / {}^5 D_0$, $^5 PF_2$, $^5 DG_3$, $^5 FH_4$;
\item[(iii)] Three coupled channels: $^1 P_1 / {}^5 PF_1$;
\item[(iv)] Four coupled channels:   $^1 D_2 / {}^5 SDG_2$.
\end{itemize}

\begin{table*}[t]
\begin{ruledtabular}
\begin{tabular}{@{\extracolsep{\fill}}lccccccc}
$n_\mathrm{ch}$ & $L$ (units of $a$) & $R_I$ (units of $a$) & $R_O$ (units of $a$) & $R_W$ (units of $a$) & $V_0$ (MeV) & $U_0$ (MeV) & $n_\mathrm{eig}$ \\[4pt]
\hline
1 & 35 & 9.02 & 12.02 & 15.02 & 0 & $-$ & 10 \\
 & 41 & 9.02 & 12.02 & 18.02 & 0 & $-$ & 10 \\
 & 47 & 9.02 & 12.02 & 21.02 & 0 & $-$ & 10 \\[4pt]
2 & 35 & 9.02 & 12.02 & 15.02 & 0 & 20 & 15 \\
 & 41 & 9.02 & 12.02 & 18.02 & 0 & 20 & 15 \\[4pt]
3 & 35 & 9.02 & 12.02 & 15.02 & 0 & 10 & 70 \\
4 & 35 & 9.02 & 12.02 & 15.02 & 0 & 5 & 110 
\end{tabular}
\end{ruledtabular}
\caption{Parameters for the lattice calculation depending on the number of coupled scattering channels ($n_\mathrm{ch}$): lattice length $L$, interval $[R_I, R_O]$ for fitting wave functions, spherical wall radius $R_W$, coefficient $V_0$ of Gaussian auxiliary potential, coefficient $U_0$ of mixing potential and number of computed eigenvectors $n_\mathrm{eig}$. The lattice spacing has been chosen as $a = 1.9733 \; \mathrm{fm}$.
\label{tab:Param}}
\end{table*}

\begin{figure*}[tb]
\centering
\includegraphics[scale=1]{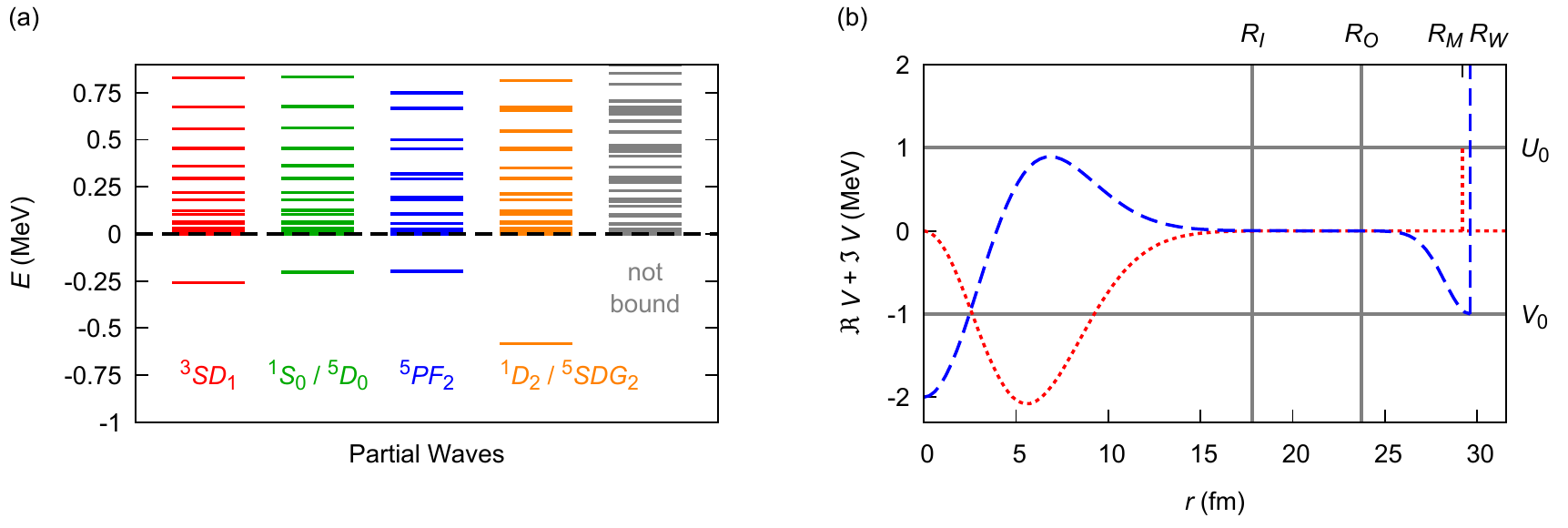}
\caption{
(a)~Spectrum of eigenenergies in the continuum for different
partial waves. The last column shows the combined spectrum in all
channels that do not have bound states. 
(b)~The employed potentials in the
  $^3 SD_1$ partial waves (adopted from Ref.~\cite{Lu:2015riz}). The blue
  dashed and red dotted lines show the diagonal $^3 D_1$-wave
  element and the off-diagonal $^3 SD_1$-wave element in the
  $2 \times 2$ potential matrix, respectively. $V_0$ and $U_0$ give the
  strengths of the auxiliary and mixing potentials, respectively,
  $R_M$ shows an approximate position of the mixing potential
  while  $R_W$ is the spherical wall radius. The wave
  function is fitted in the interval $[R_I, R_O]$.
\label{fig:Radii}}
\end{figure*}
We further emphasize that the considered potential possesses four
bound states in the $^3 SD_1$,
$^1 S_0 / {}^5 D_0$,
$^5 PF_2$ and $^1 D_2 / {}^5 SDG_2$ channels with the binding energies of
$-0.258 \; \mathrm{MeV}$, $-0.204 \; \mathrm{MeV}$, $-0.198 \; \mathrm{MeV}$, and $-0.583 \; \mathrm{MeV}$, respectively, as
visualized in Fig.~\ref{fig:Radii}(a). The binding energies
have been obtained by computing the eigenvalues of the momentum-space
Hamiltonian in the infinite-volume continuum.

\begin{figure*}[tb]
\centering
\includegraphics[scale=1]{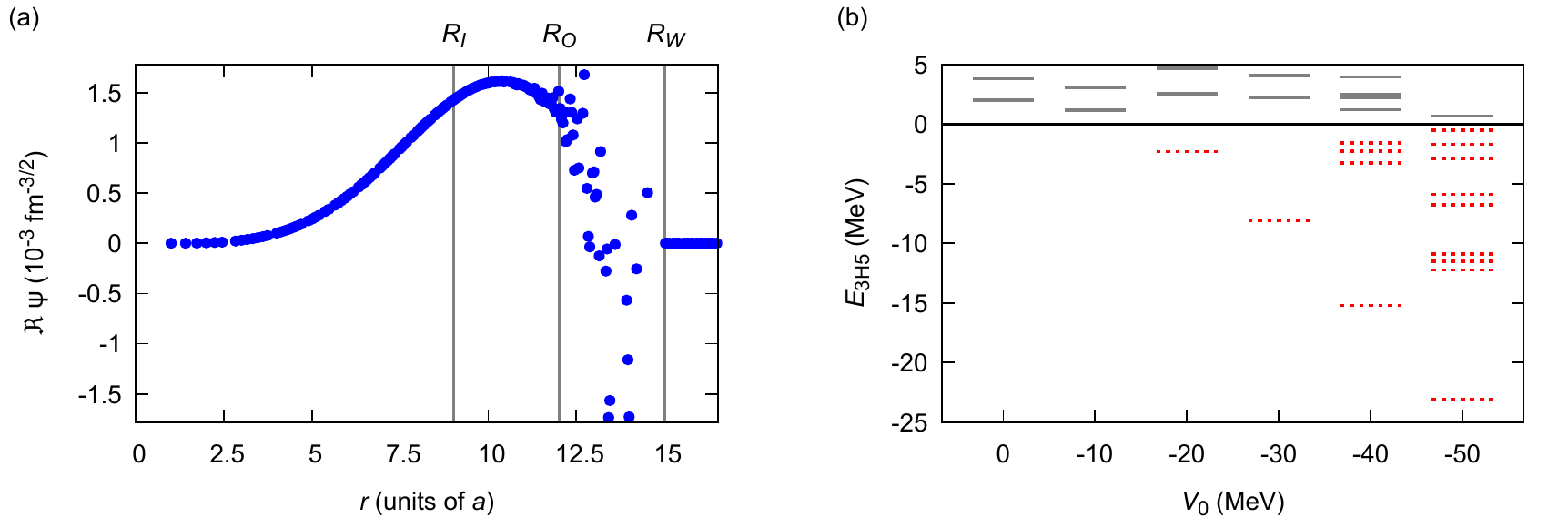}
\caption{Effects of the Gaussian auxiliary potential. 
(a)~Wave function in the $^3 H_5$ partial
wave distorted by the auxiliary potential of the strength $V_0 = -40
\; \mathrm{MeV}$, which corresponds to the outlying data point at $p
\simeq 65 \; \mathrm{MeV}$ in the last plot in Fig.~\ref{fig:1Ch}.
(b)~Eigenenergies on the lattice in the
$^3 H_5$ partial wave as functions of the
strength $V_0$ of the auxiliary potential. Large negative values of $V_0$ lead to additional bound states shown as red dotted lines.
\label{fig:Vaux}}
\end{figure*}

On the lattice, the eigenvectors of the Hamiltonian corresponding to
the lowest positive eigenenergies have been used as radial wave
functions. The parameters used in our calculations, which have partly been adopted from 
Ref.~\cite{Lu:2015riz}, can be found in Table~\ref{tab:Param} while Fig.~\ref{fig:Radii}(b) visualizes the different
contributions to the potential on the lattice.  
Notice that the Gaussian auxiliary potential can distort the wave function and may
even generate additional bound states, see Fig.~\ref{fig:Vaux}.
This may result in the appearance of the outlying points in the
calculated phase shifts or mixing angles as will be discussed below. 
For this reason, instead of imposing  the auxiliary
potential, we have actually varied the lattice size to generate  data at low
momenta. 

\begin{figure*}[tb]
\centering
\includegraphics[scale=1]{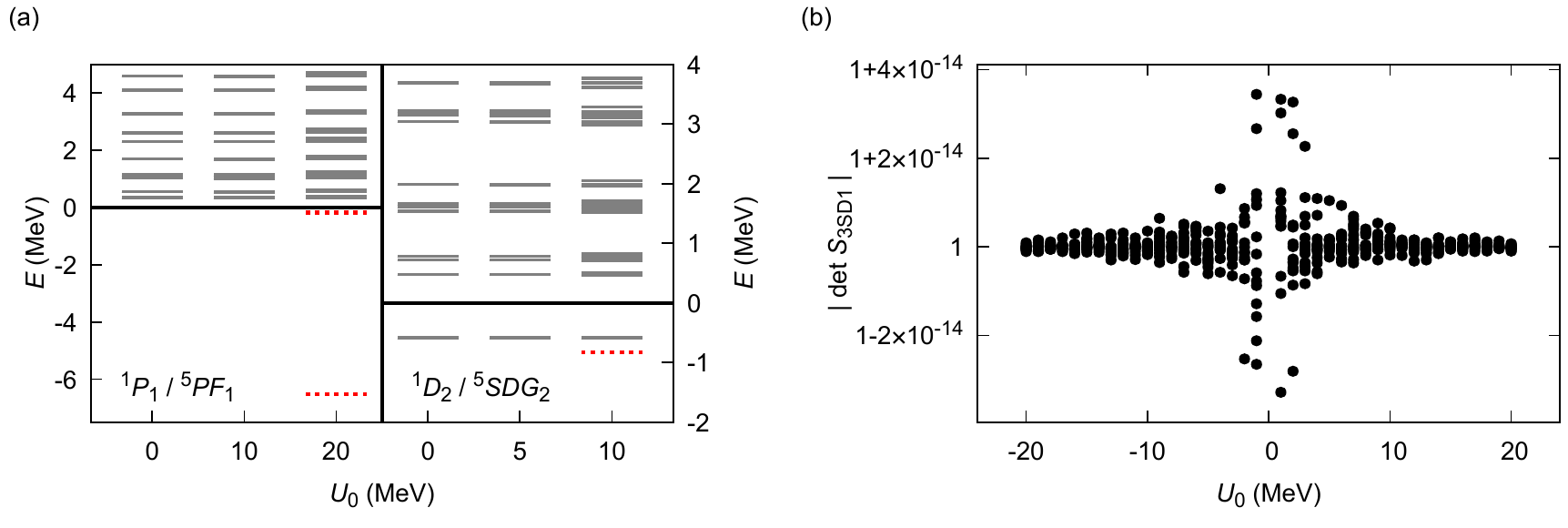}
\caption{Effects of the mixing potential.
(a)~Eigenenergies on the lattice in the
$^1 P_1 / {}^5 PF_1$ 
($^1 D_2 / {}^5 SDG_2$) partial
waves for the choices of $U_0 = 0$, $10$ and $20$~MeV ($U_0 = 0$, $5$
and $10$~MeV). Large values of the strength $U_0$ of the mixing
potential cause the appearance of additional bound states shown by red dotted lines.
(b)~Absolute values of the determinant of the $S$~matrix for all
eigenenergies in the $^3 SD_1$ partial wave as
functions of the strength $U_0$ of the mixing potential. Small in
magnitude values of $U_0$ lead to numerical instabilities resulting in
a nonunitary $S$~matrix, i.e.,~$\vert \det S \, \vert \neq 1$.
\label{fig:Vmix}}
\end{figure*}

\begin{figure*}[tb]
\centering
\includegraphics[scale=1]{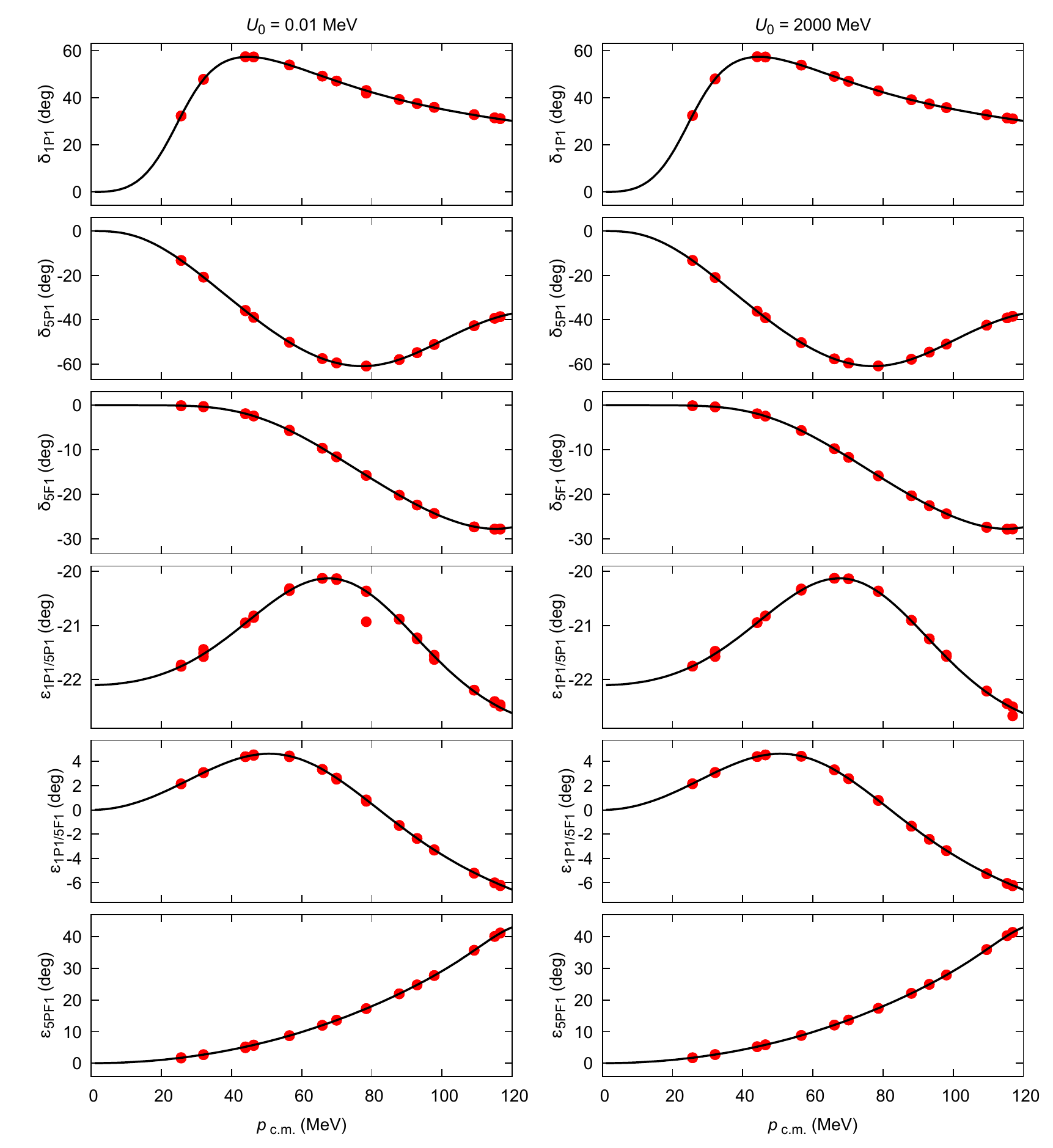}
\caption{Effects of the mixing potential with different coefficients $U_0$ on the phase shifts and mixing angles for the $^1 P_1 / {}^5 PF_1$~wave (black solid line: continuum; red points: lattice). For small coefficients (such as $U_0 = 0.01 \; \mathrm{MeV}$ in the left column) and for large coefficients (such as $U_0 = 2000 \; \mathrm{MeV}$ in the right column), outlying data points appear in the plot for the $^1 P_1 / {}^5 P_1$-wave mixing angle.
\label{fig:3ChVarU0}}
\end{figure*}

The choice of the mixing potential requires some care, too. In
particular, as shown in Table~\ref{tab:Param}, the numerical value of
the strength $U_0$ has been decreased for three and four channels in
order  to avoid the appearance of additional bound states as
visualized in Fig.~\ref{fig:Vmix}(a). On the other
hand, choosing too small in magnitude values of $U_0$ leads to
numerical instabilities causing a violation of unitarity in the
calculated $S$~matrix, see Fig.~\ref{fig:Vmix}(b). 
For large or small values of the coefficient $U_0$, outlying
data points may appear in the plots of the phase shifts and mixing angles,
as illustrated in Fig.~\ref{fig:3ChVarU0}. However, the outlying points for
small coefficients are only caused by numerical round-off errors.
We have verified that they can thus
be removed using numbers with higher precision. Therefore, any value
of $U_0$ should be suitable as long as no additional bound states are
produced by the mixing potential.

\begin{figure*}[tb]
\centering
\includegraphics[scale=1]{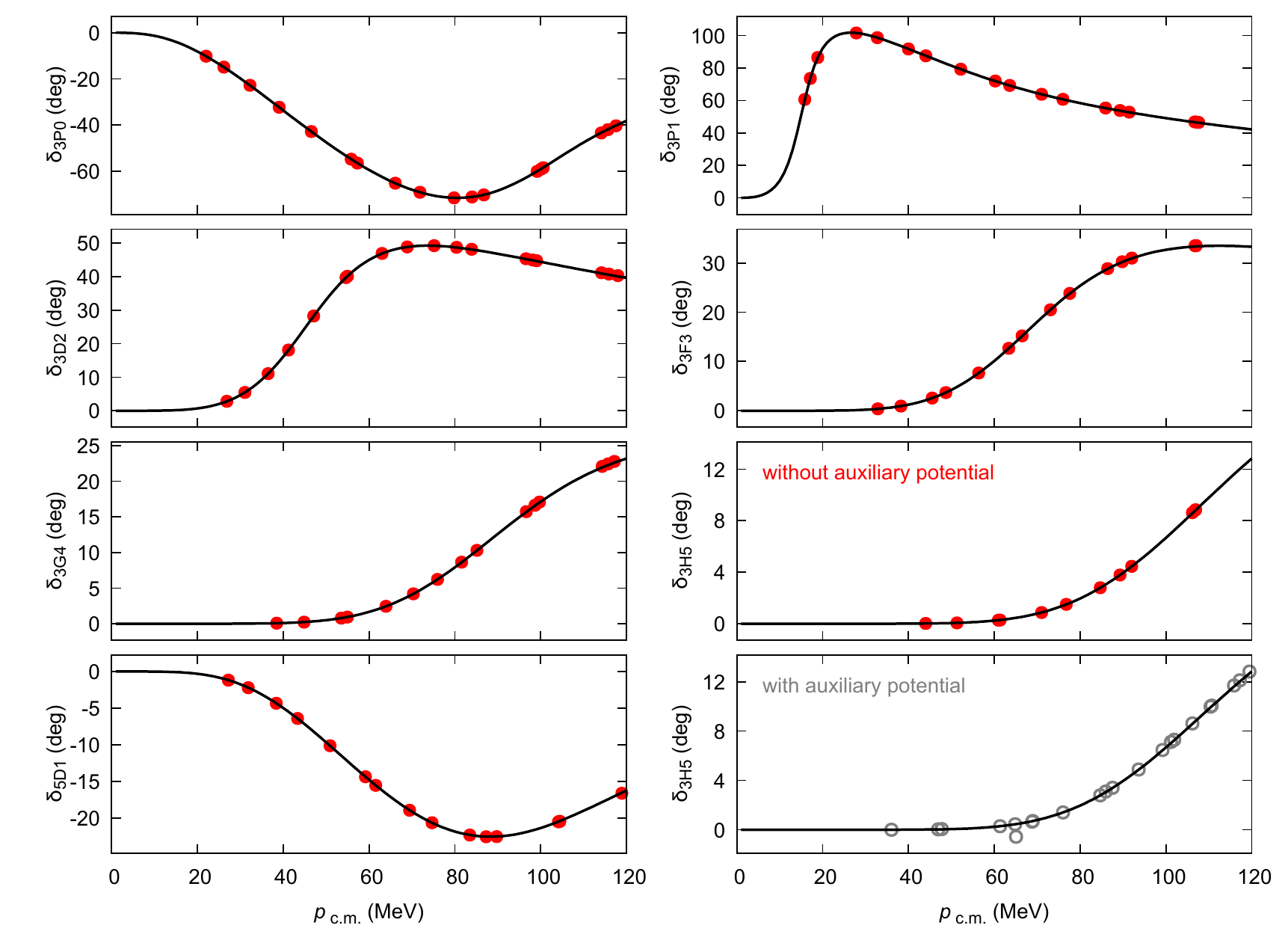}
\caption{Phase shifts in the uncoupled channels (black solid line:
  continuum; red points and gray circles: lattice). The last plot
  shows the $^3 H_5$-wave phase shift
  obtained for the auxiliary potential
  with $V_0 = 0 \; \mathrm{MeV}$, $-10 \; \mathrm{MeV}$, $\dots$, $-50
  \; \mathrm{MeV}$ at the lattice length $L = 35 a$. The outlying
  data point at $p \simeq 65 \; \mathrm{MeV}$ corresponds to the
  value of $V_0 = -40 \; \mathrm{MeV}$.
\label{fig:1Ch}}
\end{figure*}

\begin{figure*}[tb]
\centering
\includegraphics[scale=1]{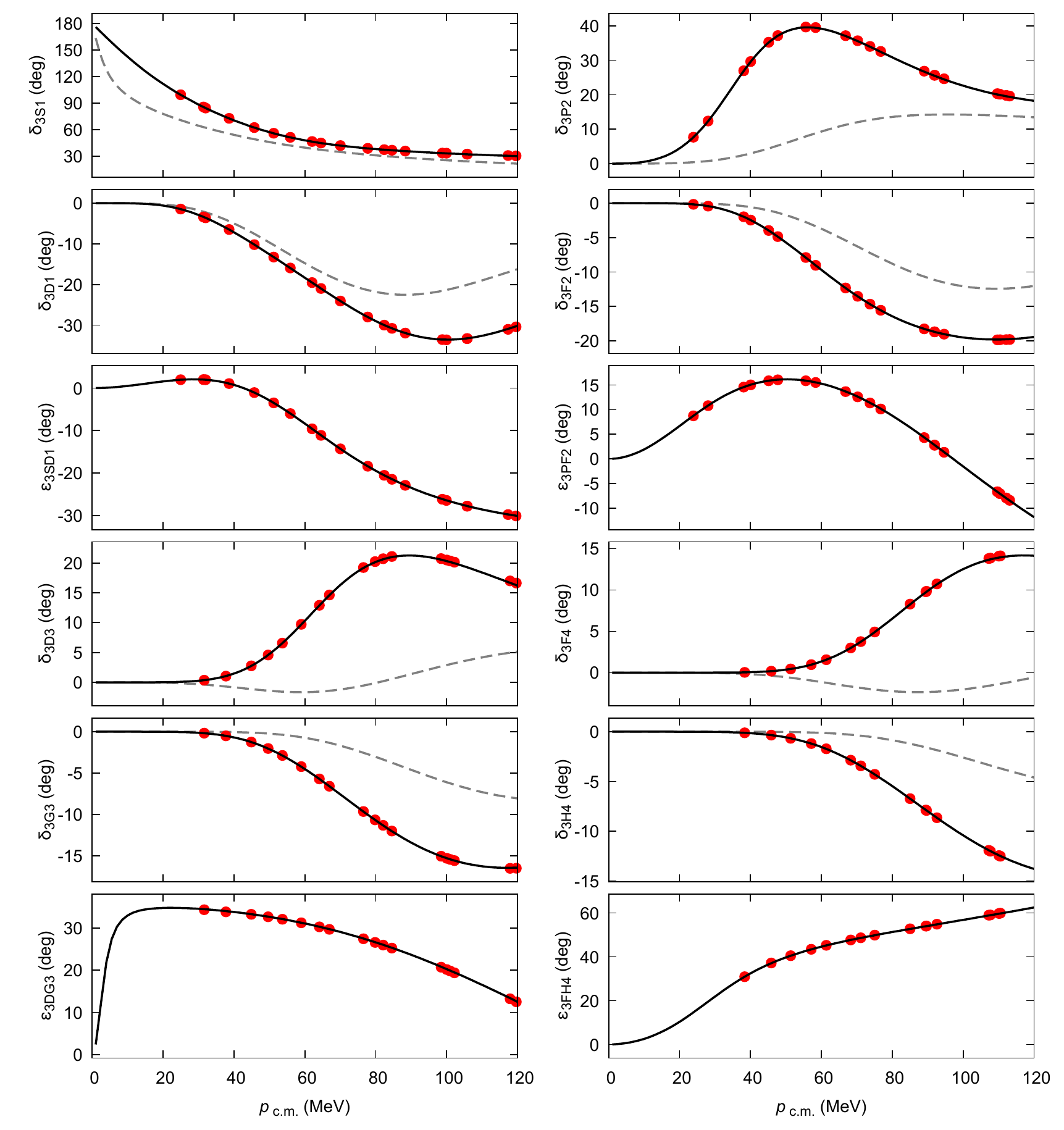}
\caption{Phase shifts and mixing angles in the pairs of coupled channels for spin $s = 1$ (black solid line: continuum; red points: lattice; gray dashed line: continuum results without channel mixing).
\label{fig:2ChS1}}
\end{figure*}

\begin{figure*}[tb]
\centering
\includegraphics[scale=1]{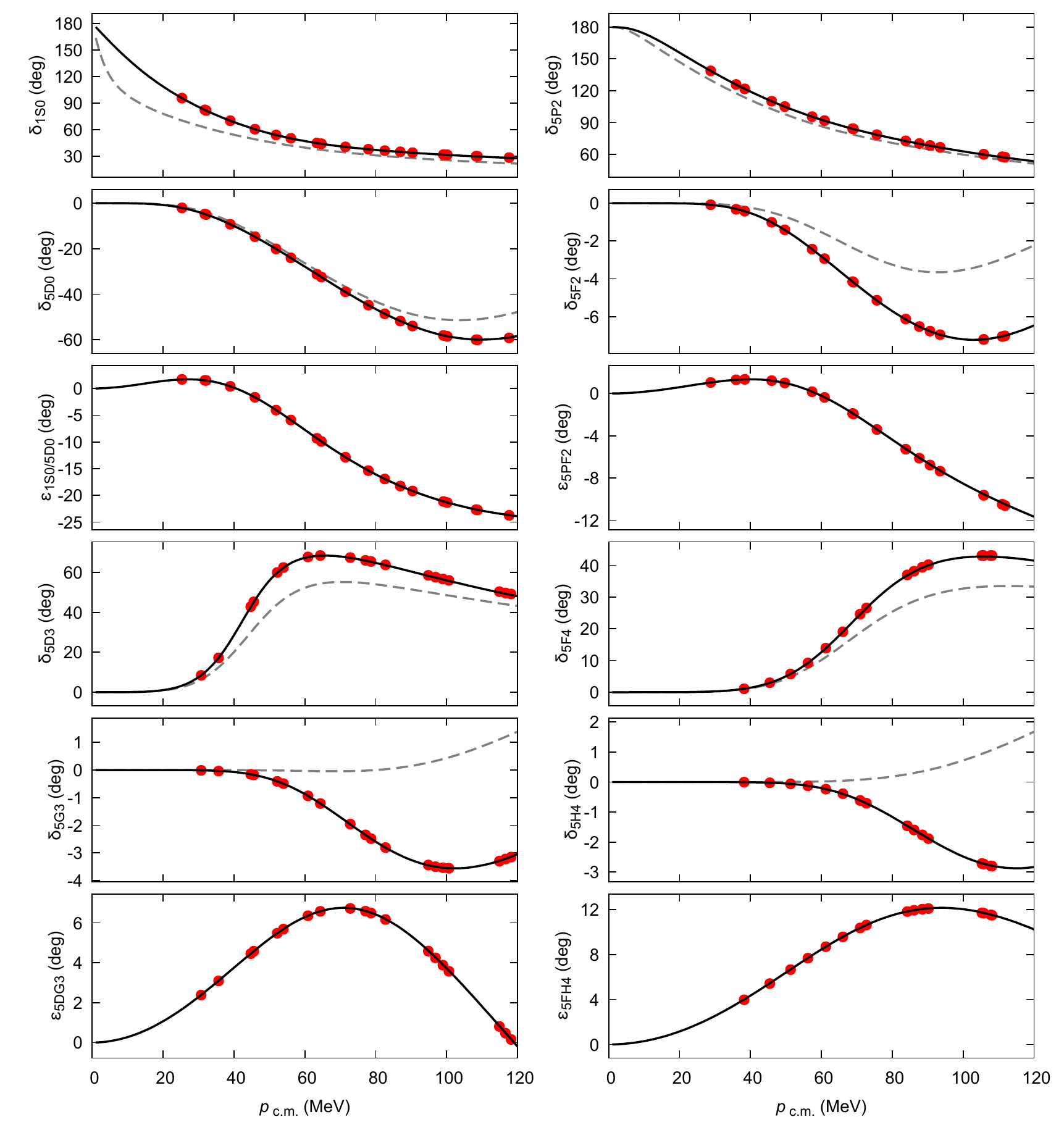}
\caption{Phase shifts and mixing angles in the pairs of coupled channels for spin $s = 0, 2$ (black solid line: continuum; red points: lattice; gray dashed line: continuum results without channel mixing).
\label{fig:2ChS02}}
\end{figure*}

\begin{figure*}[tb]
\centering
\includegraphics[scale=1]{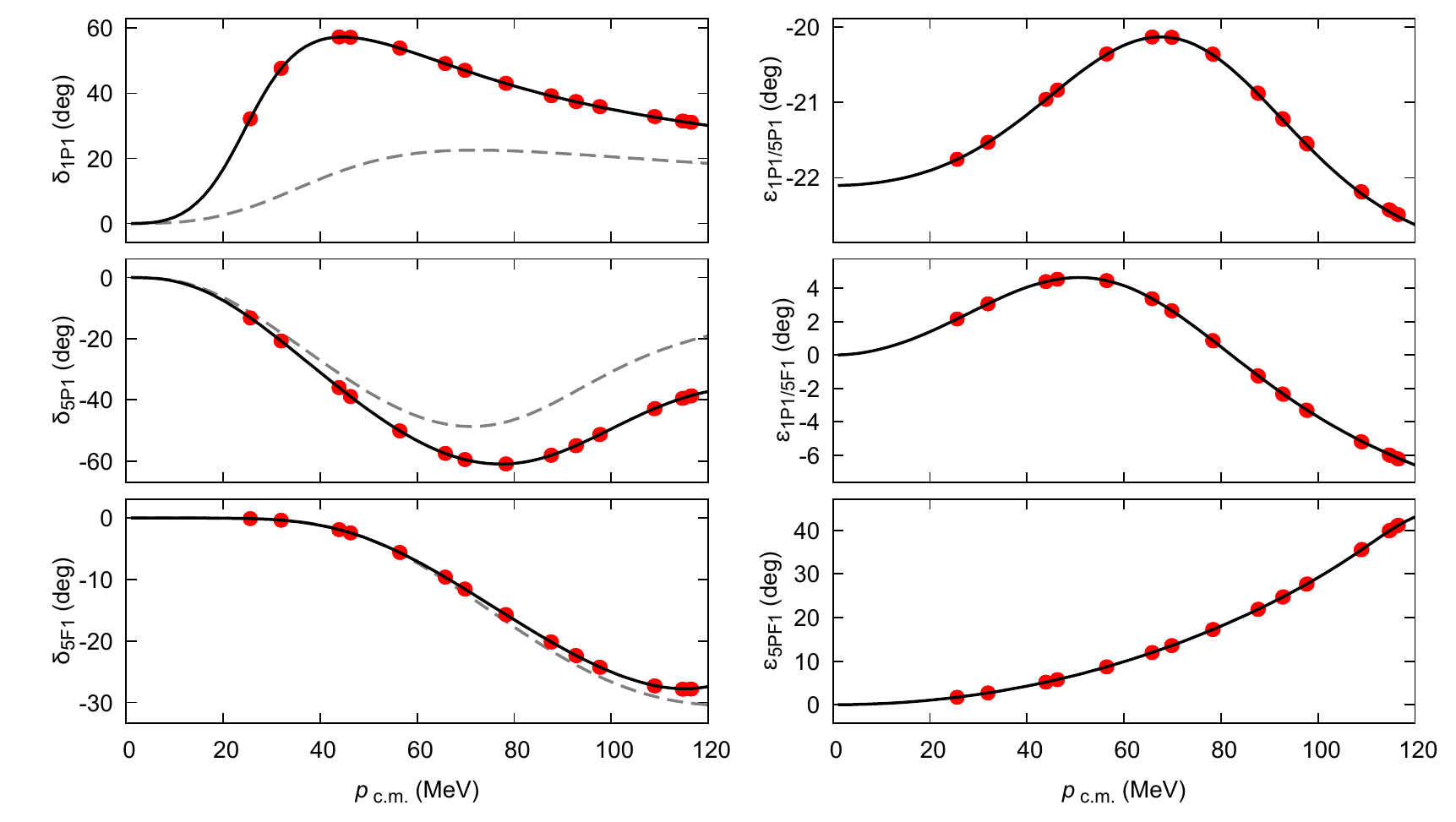}
\caption{Phase shifts and mixing angles for the $^1 P_1 / {}^5 PF_1$~wave (black solid line: continuum; red points: lattice; gray dashed line: continuum results without channel mixing).
\label{fig:3Ch}}
\end{figure*}

\begin{figure*}[tb]
\centering
\includegraphics[scale=1]{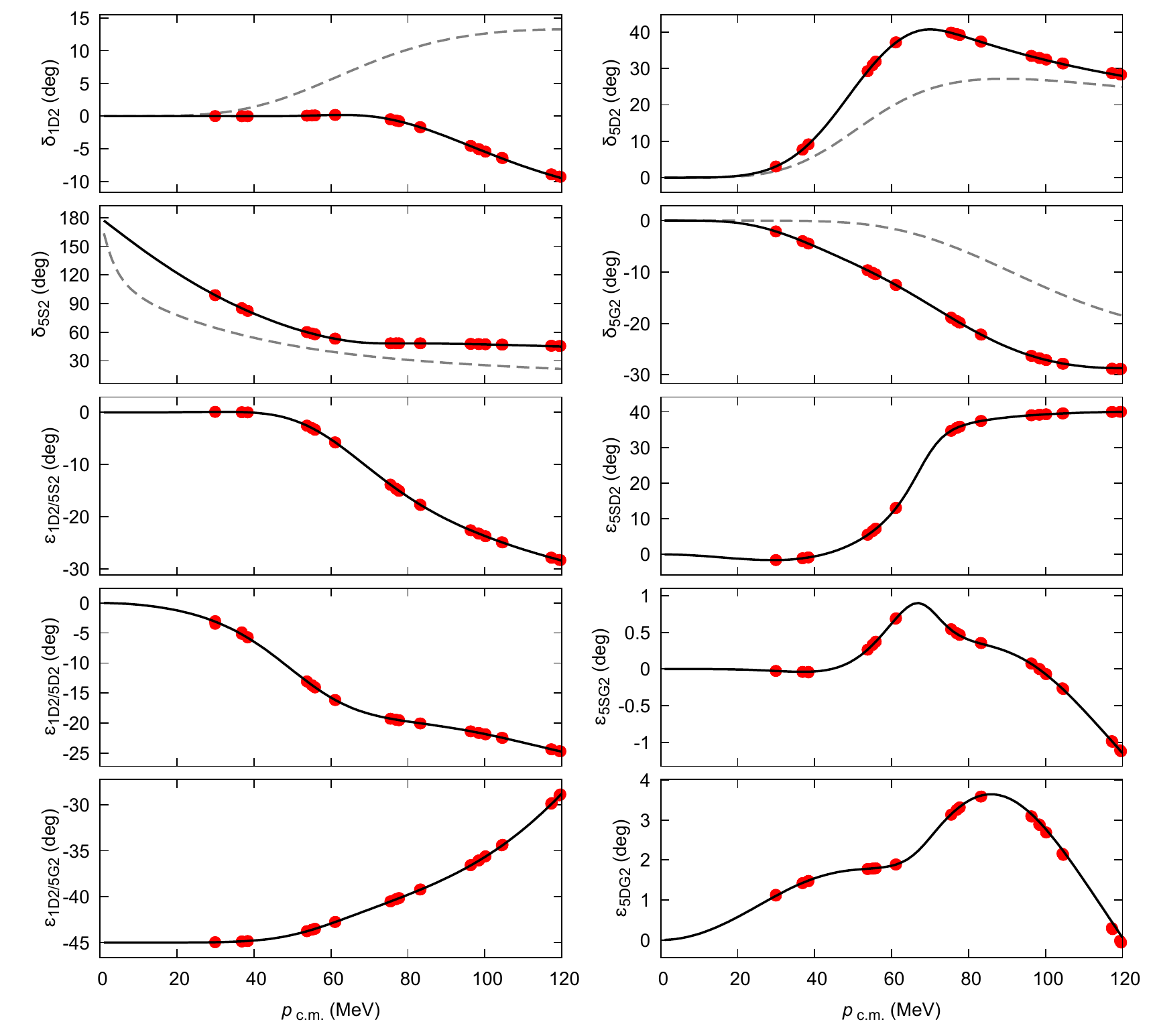}
\caption{Phase shifts and mixing angles for the $^1 D_2 / {}^5 SDG_2$~wave (black solid line: continuum; red points: lattice; gray dashed line: continuum results without channel mixing).
\label{fig:4Ch}}
\end{figure*}

In Figs.~\ref{fig:1Ch}$-$\ref{fig:4Ch}, we show the phase shifts and mixing
angles for all considered scattering channels, which have been calculated on the lattice
using the method presented above. To benchmark our calculations, we
have also computed the scattering parameters in the continuum. This
has been achieved by
solving the radial Schr\"odinger equation for a fixed energy $E =
p^2/(2 \mu)$. The boundary conditions at $r
\simeq 0$ have to be chosen in such a way that one obtains a
sufficient number of linearly independent
solutions. Finally, the $S$~matrix can again be extracted from the wave
function by fitting spherical Hankel functions. 
As shown in Figs.~\ref{fig:1Ch}$-$\ref{fig:4Ch}, the results of the
lattice calculations are in essentially a perfect agreement with the
ones calculated using the continuum approach in the considered range
of  c.m. system momenta up to  $p =
120 \; \mathrm{MeV}$.  The cutoff momentum associated with the lattice
spacing of $a = (100 \; \mathrm{MeV})^{-1} = 1.9733 \; \mathrm{fm}$
employed in our analysis is $\Lambda_\mathrm{latt}\sim \pi/a \simeq
314$~MeV. Thus, the lattice and continuum results are expected to
agree for momenta well below $\Lambda_\mathrm{latt}$. Indeed, for c.m. system
momenta higher than $p = 120 \; \mathrm{MeV}$, the deviations between
the continuum and lattice results start to become visible. This
observation is in line with the findings of Refs.~\cite{Borasoy:2007vy,Lu:2015riz}. 

We also notice a subtlety in the extraction of phase shifts in
multichannel cases due to the fact that the eigenvalues $e^{2i\delta_1}, \dots, e^{2i\delta_n}$ of
the $S$~matrix have no predefined ordering
\cite{Blatt:1952zza}. Therefore, the multichannel continuum
calculation has been repeated without the off-diagonal elements in the
potential matrix, see the gray dashed lines in
Figs.~\ref{fig:2ChS1}$-$\ref{fig:4Ch}. Then, the phase shifts in the
coupled channels have been ordered such that they are roughly
consistent with the phase shifts obtained without the coupling.
The large differences between the solid and dashed lines demonstrate
the very important role of channel mixing in the considered toy model. 
For
many-body systems where no continuum calculation is possible, the
comparison can be performed with lattice data instead. If the results
at very low momenta are available, one may also possibly identify the partial
waves from the threshold behavior of the eigenphases. 
Last but not least, we emphasize that the behavior the phase shifts in the
$^3 SD_1$, $^1 S_0 / {}^5 D_0$, $^5 PF_2$, and $^1 D_2 / {}^5 SDG_2$
channels with $\delta(p = 0)  = \pi$ is consistent with the appearance
of a single bound state in each of these channels, see Fig.~\ref{fig:Radii}(a),
in agreement with Levinson's theorem.

\section{Summary and outlook}
\label{sec4}

In this paper, we considered two-particle scattering by
solving the Schr\"odinger equation on the lattice.
A generalization of the method used in
Ref.~\cite{Lu:2015riz} for spin-$1/2$ fermions to scattering of
particles with any spin 
and an arbitrary number of coupled scattering channels has been proposed.
For the case of two spin-$1$ bosons, the proposed method was
benchmarked against the continuum approach and 
demonstrated to yield accurate and reliable results for phase shifts and mixing
angles for momenta well below the lattice cutoff. Our study opens the
way to perform \textit{ab initio} chiral EFT calculations in the
four-nucleon continuum and to access nuclear reactions on the lattice
using the adiabatic projection method. Work along these lines is in
progress.

\begin{acknowledgments}
We are grateful  to Ning Li and Bing-Nan Lu for helpful discussions and to
Ulf-G.~Mei{\ss}ner and Xiu-Lei Ren for useful comments on the manuscript.
We also thank Ning Li for sharing his nucleon-nucleon scattering code.
This work was supported by
DFG (SFB/TR 110, ``Symmetries and the Emergence of Structure in
QCD''), the BMBF  (Grant No. 05P2015) and the U.S. Department of
Energy (Grants No. DE-$ $SC0018638 and No. DE-AC52-06NA25396).
\end{acknowledgments}



\begin{thebibliography}{99}

  \bibitem{Tanabashi:2018oca} 
  M.~Tanabashi \textit{et al.} (Particle Data Group),
  Phys.\ Rev.\ D \textbf{98}, 030001 (2018).


\bibitem{Lage:2009zv}
  M.~Lage, Ulf-G.~Mei{\ss}ner, and A.~Rusetsky,
  Phys.\ Lett.\ B \textbf{681}, 439 (2009)
  [\eprint{arXiv:0905.0069 [hep-lat]}].

\bibitem{Bernard:2010fp}
  V.~Bernard, M.~Lage, Ulf-G.~Mei{\ss}ner, and A.~Rusetsky,
  J. High Energy Phys. 01 (2011) 019
  [\eprint{arXiv:1010.6018 [hep-lat]}]. 

  \bibitem{Hansen:2012tf} 
  M.~T.~Hansen and S.~R.~Sharpe,
  Phys.\ Rev.\ D \textbf{86}, 016007 (2012)
  [\eprint{arXiv:1204.0826 [hep-lat]}].

\bibitem{Briceno:2012rv} 
  R.~A.~Briceno and Z.~Davoudi,
  Phys.\ Rev.\ D \textbf{87}, 094507 (2013)
  [\eprint{arXiv:1212.3398 [hep-lat]}].

\bibitem{Hammer:2017uqm} 
  H.-W.~Hammer, J.-Y.~Pang, and A.~Rusetsky,
  J. High Energy Phys. 09 (2017), 109
  [\eprint{arXiv:1706.07700 [hep-lat]}].

  \bibitem{Hammer:2017kms} 
  H.-W.~Hammer, J.-Y.~Pang, and A.~Rusetsky,
  J. High Energy Phys. 10 (2017), 115
  [\eprint{arXiv:1707.02176 [hep-lat]}].

\bibitem{Briceno:2017tce} 
  R.~A.~Brice{\~n}o, M.~T.~Hansen, and S.~R.~Sharpe,
  Phys.\ Rev.\ D \textbf{95}, 074510 (2017)
  [\eprint{arXiv:1701.07465 [hep-lat]}].
  
  \bibitem{Briceno:2017max} 
  R.~A.~Briceno, J.~J.~Dudek, and R.~D.~Young,
  Rev.\ Mod.\ Phys.\  \textbf{90}, 025001 (2018)
  [\eprint{arXiv:1706.06223 [hep-lat]}].

  
\bibitem{Epelbaum:2014efa} 
  E.~Epelbaum, H.~Krebs, and Ulf-G.~Mei{\ss}ner,
  Eur.\ Phys.\ J.\ A \textbf{51}, 53 (2015)
  [\eprint{arXiv:1412.0142 [nucl-th]}].

\bibitem{Epelbaum:2014sza} 
  E.~Epelbaum, H.~Krebs, and Ulf-G.~Mei{\ss}ner,
  Phys.\ Rev.\ Lett.\  \textbf{115}, 122301 (2015)
  [\eprint{arXiv:1412.4623 [nucl-th]}].
  
\bibitem{Entem:2014msa} 
  D.~R.~Entem, N.~Kaiser, R.~Machleidt, and Y.~Nosyk,
  Phys.\ Rev.\ C \textbf{91}, 014002 (2015)
  [\eprint{arXiv:1411.5335 [nucl-th]}].
  
\bibitem{Entem:2017gor} 
  D.~R.~Entem, R.~Machleidt, and Y.~Nosyk,
  Phys.\ Rev.\ C \textbf{96}, 024004 (2017)
  [\eprint{arXiv:1703.05454 [nucl-th]}].
  
\bibitem{Reinert:2017usi} 
  P.~Reinert, H.~Krebs, and E.~Epelbaum,
  Eur.\ Phys.\ J.\ A \textbf{54}, 86 (2018)
  [\eprint{arXiv:1711.08821 [nucl-th]}].

  
\bibitem{Ishikawa:2007zz} 
  S.~Ishikawa and M.~R.~Robilotta,
  Phys.\ Rev.\ C \textbf{76}, 014006 (2007)
  [\eprint{arXiv:0704.0711 [nucl-th]}].

  
\bibitem{Bernard:2007sp} 
  V.~Bernard, E.~Epelbaum, H.~Krebs, and Ulf-G.~Mei{\ss}ner,
  Phys.\ Rev.\ C \textbf{77}, 064004 (2008)
  [\eprint{arXiv:0712.1967 [nucl-th]}].

\bibitem{Bernard:2011zr} 
  V.~Bernard, E.~Epelbaum, H.~Krebs, and Ulf-G.~Mei{\ss}ner,
  Phys.\ Rev.\ C \textbf{84}, 054001 (2011)
  [\eprint{arXiv:1108.3816 [nucl-th]}].

  \bibitem{Epelbaum:2007us} 
  E.~Epelbaum,
  Eur.\ Phys.\ J.\ A \textbf{34}, 197 (2007)
  [\eprint{arXiv:0710.4250 [nucl-th]}].

\bibitem{Krebs:2012yv} 
  H.~Krebs, A.~Gasparyan, and E.~Epelbaum,
  Phys.\ Rev.\ C \textbf{85}, 054006 (2012)
  [\eprint{arXiv:1203.0067 [nucl-th]}].

\bibitem{Krebs:2013kha} 
  H.~Krebs, A.~Gasparyan, and E.~Epelbaum,
  Phys.\ Rev.\ C \textbf{87}, 054007 (2013)
  [\eprint{arXiv:1302.2872 [nucl-th]}].

\bibitem{Epelbaum:2014sea} 
  E.~Epelbaum, A.~M.~Gasparyan, H.~Krebs, and C.~Schat,
  Eur.\ Phys.\ J.\ A \textbf{51}, 26 (2015)
  [\eprint{arXiv:1411.3612 [nucl-th]}].

\bibitem{Girlanda:2011fh} 
  L.~Girlanda, A.~Kievsky, and M.~Viviani,
  Phys.\ Rev.\ C \textbf{84}, 014001 (2011)
  [\eprint{arXiv:1102.4799 [nucl-th]}].

\bibitem{Epelbaum:2008ga} 
  E.~Epelbaum, H.~W.~Hammer, and Ulf-G.~Mei{\ss}ner,
  Rev.\ Mod.\ Phys.\  \textbf{81}, 1773 (2009)
  [\eprint{arXiv:0811.1338 [nucl-th]}].

\bibitem{Epelbaum:2012vx} 
  E.~Epelbaum and Ulf-G.~Mei{\ss}ner,
  Annu.\ Rev.\ Nucl.\ Part.\ Sci.\  \textbf{62}, 159 (2012)
  [\eprint{arXiv:1201.2136 [nucl-th]}].
  
\bibitem{Machleidt:2011zz} 
  R.~Machleidt and D.~R.~Entem,
  Phys.\ Rep.\  \textbf{503}, 1 (2011)
  [\eprint{arXiv:1105.2919 [nucl-th]}].
  
  
\bibitem{Gloeckle:1995jg} 
  W.~Gl\"ockle, H.~Witala, D.~Huber, H.~Kamada, and J.~Golak,
  Phys.\ Rep.\  \textbf{274}, 107 (1996).

\bibitem{Barrett:2013nh} 
  B.~R.~Barrett, P.~Navratil, and J.~P.~Vary,
  Prog.\ Part.\ Nucl.\ Phys.\  \textbf{69}, 131 (2013).

\bibitem{Hagen:2012fb} 
  G.~Hagen, M.~Hjorth-Jensen, G.~R.~Jansen, R.~Machleidt, and T.~Papenbrock,
  Phys.\ Rev.\ Lett.\  \textbf{109}, 032502 (2012)
  [\eprint{arXiv:1204.3612 [nucl-th]}].

\bibitem{Hergert:2012nb} 
  H.~Hergert, S.~K.~Bogner, S.~Binder, A.~Calci, J.~Langhammer, R.~Roth, and A.~Schwenk,
  Phys.\ Rev.\ C \textbf{87}, 034307 (2013)
  [\eprint{arXiv:1212.1190 [nucl-th]}].

\bibitem{Soma:2012zd} 
  V.~Soma, C.~Barbieri, and T.~Duguet,
  Phys.\ Rev.\ C \textbf{87}, 011303(R) (2013)
  [\eprint{arXiv:1208.2472 [nucl-th]}].

\bibitem{Lovato:2013cua} 
  A.~Lovato, S.~Gandolfi, R.~Butler, J.~Carlson, E.~Lusk, S.~C.~Pieper, and R.~Schiavilla,
  Phys.\ Rev.\ Lett.\  \textbf{111}, 092501 (2013)
  [\eprint{arXiv:1305.6959 [nucl-th]}].
  
  \bibitem{Lee:2008fa} 
  D.~Lee,
  Prog.\ Part.\ Nucl.\ Phys.\  \textbf{63}, 117 (2009)
  [\eprint{arXiv:0804.3501 [nucl-th]}].

  \bibitem{Lee:2016fhn} 
  D.~Lee,
  Lect.\ Notes Phys.\  \textbf{936}, 237 (2017)
  [\eprint{arXiv:1609.00421 [nucl-th]}].

\bibitem{UGMlectures}
  T.~A.~L\"ahde and Ulf-G.~Mei{\ss}ner, Lect. \ Notes. \ Phys. \
        \textbf{957}, 1 (2019). 
  
\bibitem{Epelbaum:2011md} 
  E.~Epelbaum, H.~Krebs, D.~Lee, and Ulf-G.~Mei{\ss}ner,
  Phys.\ Rev.\ Lett.\  \textbf{106}, 192501 (2011)
  [\eprint{arXiv:1101.2547 [nucl-th]}].

\bibitem{Epelbaum:2012qn} 
  E.~Epelbaum, H.~Krebs, T.~A.~L\"ahde, D.~Lee, and Ulf-G.~Mei{\ss}ner,
  Phys.\ Rev.\ Lett.\  \textbf{109}, 252501 (2012)
  [\eprint{arXiv:1208.1328 [nucl-th]}].

\bibitem{Epelbaum:2013paa} 
  E.~Epelbaum, H.~Krebs, T.~A.~L\"ahde, D.~Lee, Ulf-G.~Mei{\ss}ner, and G.~Rupak,
  Phys.\ Rev.\ Lett.\  \textbf{112}, 102501 (2014)
  [\eprint{arXiv:1312.7703 [nucl-th]}].
  
 
  \bibitem{Freer:2017gip} 
  M.~Freer, H.~Horiuchi, Y.~Kanada-En'yo, D.~Lee, and Ulf-G.~Mei{\ss}ner,
  Rev.\ Mod.\ Phys.\  \textbf{90}, 035004 (2018)
  [\eprint{arXiv:1705.06192 [nucl-th]}].

  
\bibitem{Borasoy:2007vk} 
  B.~Borasoy, E.~Epelbaum, H.~Krebs, D.~Lee, and Ulf-G.~Mei{\ss}ner,
  Eur.\ Phys.\ J.\ A \textbf{35}, 357 (2008)
  [\eprint{arXiv:0712.2993 [nucl-th]}].

\bibitem{Epelbaum:2009zsa} 
  E.~Epelbaum, H.~Krebs, D.~Lee, and Ulf-G.~Mei{\ss}ner,
  Eur.\ Phys.\ J.\ A \textbf{41}, 125 (2009)
  [\eprint{arXiv:0903.1666 [nucl-th]}].

\bibitem{Lahde:2013uqa} 
  T.~A.~L\"ahde, E.~Epelbaum, H.~Krebs, D.~Lee, Ulf-G.~Mei{\ss}ner, and G.~Rupak,
  Phys.\ Lett.\ B \textbf{732}, 110 (2014)
  [\eprint{arXiv:1311.0477 [nucl-th]}].

\bibitem{Li:2018ymw} 
  N.~Li, S.~Elhatisari, E.~Epelbaum, D.~Lee, B.~N.~Lu, and Ulf-G.~Mei{\ss}ner,
  Phys.\ Rev.\ C \textbf{98}, 044002 (2018)
  [\eprint{arXiv:1806.07994 [nucl-th]}].

\bibitem{Epelbaum:2012iu} 
  E.~Epelbaum, H.~Krebs, T.~A.~L\"ahde, D.~Lee, and Ulf-G.~Mei{\ss}ner,
  Phys.\ Rev.\ Lett.\  \textbf{110}, 112502 (2013)
  [\eprint{arXiv:1212.4181 [nucl-th]}].

\bibitem{Epelbaum:2013wla} 
  E.~Epelbaum, H.~Krebs, T.~A.~L\"ahde, D.~Lee, and Ulf-G.~Mei{\ss}ner,
  Eur.\ Phys.\ J.\ A \textbf{49}, 82 (2013)
  [\eprint{arXiv:1303.4856 [nucl-th]}].

\bibitem{Meissner:2014pma} 
  Ulf-G.~Mei{\ss}ner,
  Sci.\ Bull.\  \textbf{60}, 43 (2015)
  [\eprint{arXiv:1409.2959 [hep-th]}].
  
\bibitem{Elhatisari:2017eno} 
  S.~Elhatisari, E.~Epelbaum, H.~Krebs, T.~A.~L\"ahde, D.~Lee, N.~Li, B.~N.~Lu, Ulf-G.~Mei{\ss}ner, and G.~Rupak,
  Phys.\ Rev.\ Lett.\  \textbf{119}, 222505 (2017)
  [\eprint{arXiv:1702.05177 [nucl-th]}].

\bibitem{Lu:2018bat} 
  B.~N.~Lu, N.~Li, S.~Elhatisari, D.~Lee, E.~Epelbaum, and Ulf-G.~Mei{\ss}ner,
  Phys.\ Lett.\ B \textbf{797}, 134863 (2019)
  [\eprint{arXiv:1812.10928 [nucl-th]}].

\bibitem{Alarcon:2017zcv} 
  J.~M.~Alarcon, D.~Du, N.~Klein, T.~A.~L\"ahde, D.~Lee, N.~Li, B.~N.~Lu, T.~Luu, and Ulf-G.~Mei{\ss}ner,
  Eur.\ Phys.\ J.\ A \textbf{53}, 83 (2017)
  [\eprint{arXiv:1702.05319 [nucl-th]}].
  
  \bibitem{Klein:2018iqa} 
  N.~Klein, D.~Lee, and Ulf-G.~Mei{\ss}ner,
  Eur.\ Phys.\ J.\ A \textbf{54}, 233 (2018)
  [\eprint{arXiv:1807.04234 [hep-lat]}].

\bibitem{Li:2019ldq} 
  N.~Li, S.~Elhatisari, E.~Epelbaum, D.~Lee, B.~Lu, and Ulf-G.~Mei{\ss}ner,
  Phys.\ Rev.\ C \textbf{99}, 064001 (2019)
  [\eprint{arXiv:1902.01295 [nucl-th]}].

\bibitem{Briceno:2014oea} 
  R.~A.~Briceno,
  Phys.\ Rev.\ D \textbf{89}, 074507 (2014)
  [\eprint{arXiv:1401.3312 [hep-lat]}].

\bibitem{Moir:2016srx} 
  G.~Moir, M.~Peardon, S.~M.~Ryan, C.~E.~Thomas, and D.~J.~Wilson,
  J. High Energy Phys. 10 (2016) 011
  [\eprint{arXiv:1607.07093 [hep-lat]}].

\bibitem{Briceno:2017qmb} 
  R.~A.~Briceno, J.~J.~Dudek, R.~G.~Edwards, and D.~J.~Wilson,
  Phys.\ Rev.\ D \textbf{97}, 054513 (2018)
  [\eprint{arXiv:1708.06667 [hep-lat]}].

\bibitem{Woss:2018irj} 
  A.~J.~Woss, C.~E.~Thomas, J.~J.~Dudek, R.~G.~Edwards, and D.~J.~Wilson,
  J. High Energy Phys. 07 (2018) 043
  [\eprint{arXiv:1802.05580 [hep-lat]}].

\bibitem{Woss:2019hse} 
  A.~J.~Woss, C.~E.~Thomas, J.~J.~Dudek, R.~G.~Edwards, and D.~J.~Wilson,
  Phys.\ Rev.\ D \textbf{100}, 054506 (2019)
  [\eprint{arXiv:1904.04136 [hep-lat]}].

\bibitem{Carlson:1984zz} 
  J.~Carlson, V.~R.~Pandharipande, and R.~B.~Wiringa,
  Nucl.\ Phys.\ A \textbf{424}, 47 (1984).

\bibitem{Borasoy:2007vy} 
  B.~Borasoy, E.~Epelbaum, H.~Krebs, D.~Lee, and Ulf-G.~Mei{\ss}ner,
  Eur.\ Phys.\ J.\ A \textbf{34}, 185 (2007)
  [\eprint{arXiv:0708.1780 [nucl-th]}].

\bibitem{Lu:2015riz} 
  B.-N.~Lu, T.~A.~L\"ahde, D.~Lee, and Ulf-G.~Mei{\ss}ner,
  Phys.\ Lett.\ B \textbf{760}, 309 (2016)
  [\eprint{arXiv:1506.05652 [nucl-th]}].

\bibitem{Rokash:2015hra} 
  A.~Rokash, M.~Pine, S.~Elhatisari, D.~Lee, E.~Epelbaum, and H.~Krebs,
  Phys.\ Rev.\ C \textbf{92}, 054612 (2015)
  [\eprint{arXiv:1505.02967 [nucl-th]}].

  \bibitem{Elhatisari:2016hby} 
  S.~Elhatisari, D.~Lee, Ulf-G.~Mei{\ss}ner, and G.~Rupak,
  Eur.\ Phys.\ J.\ A \textbf{52}, 174 (2016)
  [\eprint{arXiv:1603.02333 [nucl-th]}].
  
\bibitem{Elhatisari:2015iga} 
  S.~Elhatisari, D.~Lee, G.~Rupak, E.~Epelbaum, H.~Krebs, T.~A.~L\"ahde, T.~Luu, and Ulf-G.~Mei{\ss}ner,
  Nature (London) \textbf{528}, 111 (2015)
  [\eprint{arXiv:1506.03513 [nucl-th]}].

  \bibitem{Blatt:1952zza} 
  J.~M.~Blatt and L.~C.~Biedenharn,
  Phys.\ Rev.\  \textbf{86}, 399 (1952).

  
\end{thebibliography}
\end{document}